# Stellar Aspects of Habitability – Characterizing Target Stars for Terrestrial Planet Search Missions

Lisa Kaltenegger[1], Carlos Eiroa[2], Ignasi Ribas[3], Francesco Paresce[4], Martin Leitzinger[5], Petra Odert[5], Arnold Hanslmeier[5], Malcolm Fridlund[6], Helmut Lammer[7], Charles Beichman[8], William Danchi[9], Thomas Henning[10], Tom Herbst[11], Alain Léger[12], René Liseau[13], Jonathan Lunine[14], Alan Penny[15], Andreas Quirrenbach[16], Huub Röttgering[17], Frank Selsis[18], Jean Schneider[19], Daphne Stam[20], Giovanna Tinetti[20], Glenn J. White[14,21]

[1]Harvard University/CfA, 60 Garden Street MS-20, Cambridge, MA, USA, Tel. (+1) 617-495-7158, e-mail: lkaltenegger@cfa.harvard.edu
[2]Universidad Autonoma de Madrid, Madrid, Spain: [3]Institut de Ciències de l'Espai (CSIC-IEEC), Barcelona, Spain: [4]INAF,Via del Parco Mellini 84, Rome, Italy: [5]Karl Franzens University Gray, Austria; [6]RSSD, ESA, ESTEC, Noordwijk, The Netherlands; [7]Space Research Institute, Austrian Academy of Sciences, Graz, Austria; [8]NASA ExoPlanet Science Institute, California Inst. Of Technology/JPL, USA; [9]Goddard Space Flight Center, Greenbelt, MD, USA; [10]Max-Planck Institut für Astronomie, Heidelberg, Germany; [11]Universite Paris-Sud, Orsay, France; [12]Dept. of Radio and Space Science, Chalmers University of Technology; Onsala, Sweden [13]Lunar and Planetary Laboratory, University of Arizona, USA; [14]Space Science & Technology Department, CCLRC Rutherford Appleton Laboratory, Oxfordshire, UK; [15]Landessternwarte, Heidelberg, Germany; [16]Leiden Observatory, Leiden, The Netherlands; [17]University of Bordeaux 1, Bordeaux, France ; [18]Observatoire de Paris-Meudon, LUTH, Meudon, France ; [19]SRON, Netherlands Institute for Space Research, Utrecht, The Netherlands; [20]Department of Physics and Astronomy, University College London, London, UK;[21]The Open University, Milton Keynes, UK

**Abstract**

In this paper we present and discuss the criteria for selecting potential target stars suitable for the search for Earth like planets, with a special emphasis on the stellar aspects of habitability. Missions that search for terrestrial exoplanets will explore the presence and habitability of Earth-like exoplanets around several hundred nearby stars, mainly F, G, K, and M stars. The evaluation of the list of potential target systems in order to develop mission concepts for a search for Terrestrial Exoplanets is essential. Using the Darwin All Sky Star Catalogue (DASSC), we discuss the selection criteria, configuration dependent sub-catalogues and the implication of stellar activity for habitability.

**Introduction**

About 350 extrasolar planets orbiting stars other than the Sun have already been detected during the first productive 13 years of searching – down to a few Earth masses. Most planets found are something more akin to the gas giant planets in our own solar system and significantly more massive than Earth ($> 5 M_{Earth}$ to $< 13 M_{Jupiter}$, the brown dwarf limit). This is in part due to selection effects, since the most successful methods for discovering individual planets are the radial velocity and the transit methods that work best for larger planets.

Currently fifteen exoplanets including three pulsar planets are known to have a mass (times sin i, where i is the orbital inclination, for RV planets) less than 10 $M_{Earth}$, a somewhat arbitrary boundary that distinguishes terrestrial from giant planets. Accordingly we identify masses in the range 1-10 $M_{Earth}$ as being Super Earths, likely composed of rock, ice, and liquid (Leger et al. 2009; Rivera et al. 2005; Beaulieu et al. 2006; Udry et al. 2007; Mayor et al. 2009; Bennett et al. 2008; Bouchy et al. 2009; Forveille et al. 2009; Howard et al. 2009;Wolszczan & Frail 1992), and masses greater that 10 $M_{Earth}$ as being giant planets, likely dominated by the mass of a gaseous envelope.

Especially Gl581 d, the first super-Earth within the HZ of its host star, as well as Gl581 c that is just outside of the HZ close to the inner edge of the HZ (Udry et al. 2007, Selsis et al. 2007, Kaltenegger et al. 2009) are extremely interesting objects. Nevertheless, at the time of writing no 'true terrestrial analogues', i.e. an Earth size body in the middle of the 'Habitable Zone' has been reported.

We expect planets where we will eventually find life to be small and rocky – very similar to Earth. Essentially all planned searches for such 'true terrestrial analogues', focus on this so-called 'Habitable Zone' (Kasting et al. 1993), which is very loosely defined as the region around a particular star, where one would expect stable conditions for liquid water on a planetary surface. The topic is very complex, since many factors unrelated to stellar luminosity also influence habitability 'like the planetary environment e.g. planet albedo, composition, and the amount of greenhouse gases in the atmosphere. Detailed simulations for space based missions that search for Earth-like exoplanets critically depend on correct data on the target stars.

Section one discusses target selection criteria in details, section two focuses on the characteristics of the DASSC, derives sub-star-catalogues based on different technical designs and shows that influence on the target stars that can be sampled.

Section three discusses the implication of stellar activity on habitability.

# 1. Target Star Selection Criteria

Several lists of nearby target stars have been compiled that are related to the topic of this paper, but intended for different purposes, e.g. SETI target star catalogue (Turnbull and Tarter 2003), Terrestrial Planet Finder Coronagraph (TPF-C) optimized for detection of reflected light from the planet (Brown 2006), as well as the closest stars to the Sun (e.g. Porto de Mello et al. 2006; Gray et al. 2003). Here we concentrate on the DASSC, a coherent, large, 30pc distance limited sample of target stars based on Hipparcos data (Kaltenegger et al.2009, Eiroa et al. 2003; Kaltenegger 2004). Note that the DASSC was compiled to model realistic observation scenarios and facilitate a detailed revised study of the target stars.

## 1.1 Spectral Type of the Target star
### 1.1.2. Main Sequence Stars

G stars are historically considered to be the prime targets for a search for habitable planets. However, since planets have now been found around essentially all different stellar types between class F and M, recent target star samples like the DASSC reflects that. Further, considering remotely detectable habitability, it is now generally thought that the main issue is the ability of water to remain liquid on the planet's surface. This requires a relatively dense atmosphere. There is nothing *a priori* excluding any spectral type from the catalog based on this requirement.

Especially interesting in this context are M dwarfs (see e.g. Reid et al. 2004, 2009, Henry et al. 2009), the most common stellar objects. Nevertheless, they have been questioned as possible host stars for habitable planets because of their faint luminosity, requiring a planet to be very close to the star in order to be within the HZ. This latter aspect may force the planet to have bound rotation (Dole 1964), which could lead to processes where key gases freeze out on the dark side. On the other hand, it has been demonstrated, through detailed modeling, (Joshi et al. 1997; Joshi 2003) that even rather thin atmospheres could be sustained without freezing out.

Red dwarf stars also are prone to flare activity orders of magnitude higher than anything our Sun has experienced, since the T-Tauri stage prior to the epochs when life arose on Earth. Such activity could inhibit the emergence of life and its subsequent evolution if life were not protected by a layer of water or soil. Further, the high levels of particle flow associated with strong flares could strip significant amounts of atmosphere from a terrestrial planet irrespective of the strength of the planets magnetic field. Many issues remain to be clarified, but Super-Earths have already been discovered orbiting red dwarfs close to and within the HZ (see e.g. Udry et al. 2007, Mayor et al. 2008). For a detailed discussion of habitability of Earth-like planets around host stars of different spectral type (F, G, K and M) and detectable biomarkers in their atmosphere, we refer the reader to Segura et al. (2003), Segura et al. (2005), Selsis (2002), Scalo et al. (2007). The evolution of biomarkers in Earth's atmosphere during different epochs has been investigated by Kaltenegger et al. (2007) in respect to instrument requirements.

**1.1.2 Early Type Stars**

In any kind of unbiased survey, it is necessary to be extremely careful in the selection. In the context of missions searching for worlds like our own, this has led to a number of different criteria being applied. Usually, when the goal is habitability, the studies focus on solar type stars, including or excluding M-dwarfs depending on the arguments for and against (see previous section), but essentially always excluding early type stars. The argument for this is of course the short main-sequence life time, being of the order of 500 million years for an early A-type star. The number of early type stars within the volume searchable with contemplated instruments is relatively low. Statistically, they will therefore not allow anything to be said of eventual non-detections. On the other hand, the time span for the origin of life on Earth was maybe as short as 10 million years. This means that it would be both interesting and – because of the brightness and low numbers available within the search volume – not very time consuming to include them in the survey.

**1.2 Number of Target Stars**

It is essential to have a certain number of targets stars to derive conclusions for detections or non-detections. This is a key problem when designing an instrument that can detect and study terrestrial exoplanets. Detecting several exoplanets would allow conclusions to be drawn with respect to the boundary conditions under which rocky planets are formed in the inner parts of planetary systems, the boundary conditions for formation and evolution, and the actual evolutionary pathways – at least in a rudimentary way – and finally increase the chance of being able to discuss habitability.

Assuming a 10% planet fraction similar to the observed EGP fraction, a minimum of 150 stars need to be observed to detect 15 terrestrial planets, if 10% of those are assumed to be habitable, that would lead to one or two habitable planets detected and characterized. For the proposed space telescopes, the trade off and limiting factor is the integration time for a detection of spectral features, what directly translates in a number of stars that can be observed for a certain size telescope during a mission's lifetime. Even though the frequency of extra-solar terrestrial planets can only be guessed at the moment, we believe that a sample has to consist of at least > 150 stars in order to be able to derive conclusions – particularly as what concerns non-detections.

**1.3. Metallicity and Hosts of EGP**

Studies of stars hosting extrasolar giant planets (EGPs) have demonstrated that EGPs are more frequent around stars with enhanced metallicity, but such a trend is not seen so far for stars hosting Super-Earths (Mayor et al. 2009). Metallicity estimates using existing Strömgren photometry and/or spectroscopy of a considerable fraction of the target stars for extrasolar planet search missions are consistent with field stars (Eiroa et al. 2003).

Within 30 pc, 64[1] stars are known to host EGP, some of these systems can be used to calibrate instruments - especially for EGP with effective temperature determined by Spitzer. In addition, characterizing these EGP systems should bring interesting results and allow the understanding of planetary systems with giant planets, and thus give clues to formation and composition of EGP. Several groups are working on the dynamical stability of terrestrial planets in the HZ in those systems (see e.g. Lunine 2001, Raymond et al. 2006, Sandor et al. 2007, Dvorak et al. this volume), making some of the systems targets for habitable terrestrial planet search.

**1.4 Stellar Ages**

The age of the stars, although not a primary constraint for the presence of Earth-like exoplanets, largely influences the planet's atmospheres and consequently its habitability. In general, the determination of stellar ages is a complicated problem. Fundamental age determination using radioactive element decay (as used to date rocks on Earth) is of very limited applicability in the case of stars (Cayrel et al. 2001). Thus, the use of theoretical stellar evolution models is often the only viable option, as commonly done for stellar ensembles (i.e., clusters). However, using stellar models becomes difficult for isolated stars, especially for those with masses below that of the Sun and ages of less than a few Gyr because of the inherent degeneracy (i.e., slow variation of stellar properties with age). For stars with young and intermediate ages alternative ways have already been proposed (see Mamajek et al. 2008). Some examples are the use of lithium abundance observations (only useful for very young objects), the kinematic membership in stellar moving groups and wide binary pairs, the time-dependence of activity parameters (so-called gyrochronology) and the use of asteroseismology. The latter is now emerging as one of the most promising approaches using very accurate photometric time-series data from space missions (e.g. CoRoT and Kepler) that allow for a detailed study of stellar oscillations and provide a direct link to the age of the star, potentially determinable to with a precision of 5-10% (Kjeldsen et al. 2009).

Since high-precision time-series photometry is not available for all stars in the solar neighborhood it is best for now to focus the attention on the use of gyrochonology since this can be used widely. In general, the activity level of low-mass stars decreases with time as they spin down from mass loss via magnetized winds. The indicators of chromospheric and coronal activity (emission in lines such as Mg II h&k, Ca II H&K, H$\alpha$, Ca II

---

[1] http://exoplanet.eu/ (March 2009)

IRT, and the overall flux in the X-rays and EUV) scale with the rotation period of the star (and its mass) and suffer a power-law decrease with time. Relationships of this form have been proposed by, e.g., Barnes (2007) and later revised by Mamajek & Hillenbrandt (2008). In general, the accuracy of the resulting age is best for the youngest stars (20-40%) and degrades severely beyond 1 Gyr. The overall decrease of the X-ray flux seems to be the most sensitive of the activity indicators. This is being exploited by Ribas et al. (2009) to propose an age calibration covering up to about 10 Gyr.

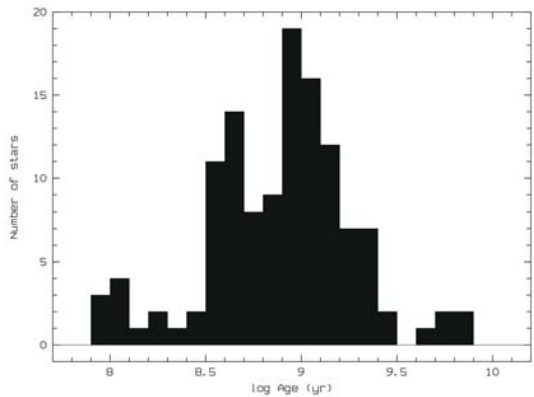

**Fig. 1: Age histogram of G and K spectral type stars within 25pc based on their ROSAT X-ray luminosity (Ribas et al. 2009).**

Fig. 1 shows an age histogram of G and K target stars within 25 pc based on their ROSAT X-ray luminosities and employing such preliminary calibration. The plot demonstrates that the closest stars constitute a good sample to study the evolution and properties of Earth-like exoplanets and their atmospheres, directly linked to their habitability (see e.g. Kaltenegger et al. 2007, Lammer et al. 2009), since a variety of ages are present.

### 1.6. The Habitable Zone

For a given planet (assuming a certain atmosphere composition and albedo) the surface temperature depends on the distance from, the luminosity of the host star and the normalized solar flux factor $S_{eff}$ that takes the wavelength dependent intensity distribution of the spectrum of different spectral classes into account. The distance d of the HZ for Earth can be calculated as (Kasting et al. 1993)

$$d = 1\ \text{AU} \times ((L/L_{sun})/S_{eff})^{0.5} \qquad (1)$$

where $S_{eff}$ is 1.90, 1.41, 1.05 and 1.05 for F, G, K and M stars respectively for the inner edge of the HZ (where runaway greenhouse occurs) and 0.46, 0.36, 0.27 and 0.27 for F, G, K and M stars respectively for the outer edge of the HZ (assuming a maximum greenhouse effect in the planet's atmosphere). These calculations were originally done for F0, G2 and K0 spectra and will be updated for all spectral sub classes (Kaltenegger & Segura 2009). The size of the HZ translates into instrument requirements such as the minimum resolution required to detect the system.

Fig. 2 shows the extent of the HZ and therefore what sub-sample we can probe for habitable planets depending on the 'IWA' or resolution of any instrument design. The extent of the HZ does not take the effect of clouds into account. The physics is, however, still very poorly understood, but the consequences should mainly move the HZ 1) toward the star if the cooling reflective properties of the clouds is stronger than their warming effect; 2) outwards if the warming effect is stronger than the cooling, or 3) potentially extending the HZ in both directions if the water clouds that form at the inner edge of the HZ have a cooling while the $CO_2$ clouds that form at the outer edge of the HZ have a warming effect (see section Biomarkers).

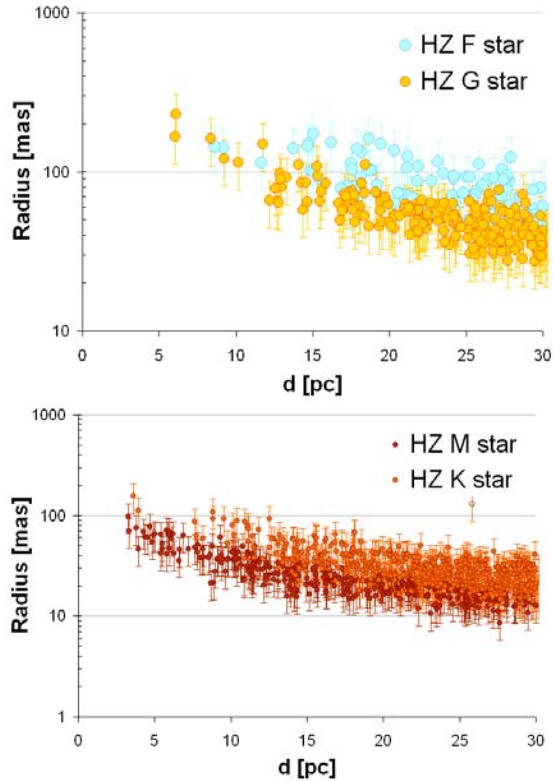

**Fig. 2 Extent of the habitable zone in mas for the DASSC.**

## 2. Darwin All Sky Star Catalogue

For the main Darwin All Sky Catalogue (DASSC)[2] (Kaltenegger et al. 2009), we selected all F, G, K and M stars within a distance of 30 pc from the Hipparcos catalogue (spectral classification of stars without an assigned spectral type is based on the B-V and V-K color index). Hipparcos data has typical parallax standard errors of about 1 milliarcsec (mas), which allows very precise distance measurements. It also includes accurate photometry (the uncertainty in B-V is typically less than 0.02mag), and proper motion data, as well as information on variability and multiplicity (Perryman et al. 1997). The DASSC combines the Hipparcos information with data from the 2MASS catalogue (Cutri et al. 2003) and the Catalogue of Components of Double and Multiple stars (CCDM (Dommanget and Nys 1994)) and the ninth catalogue of spectroscopic binary orbits (SB9 (Pourbaix et al. 2004)). Based on the combined data characteristics like luminosity, radius, mass, effective temperature and the extent of the Habitable Zone was calculated for the target star sample.

DASSC is a volume limited sample of Hipparcos stars that is magnitude-limited to about V-magnitude (Vmag) of 7.5 (Perryman et al. 1997) and includes stars to Vmag 12 (the Hipparcos catalogue is however incomplete between Vmag 7.5 and 12). A number of stars are flagged in the DASSC as being multiple. Systems that are given a multiple designation in the Hipparcos catalogue but not in the CCDM list are most probably spectroscopic binaries and introduce a bias in the data, since the derived parameters assume that the basic information is produced by a single star. Multiple systems found in the SB9 are spectroscopic binaries with orbital solutions. All multiple objects are flagged in the DASSC in order to be able to either exclude them or treat them specially when producing configuration dependent sub-catalogues for models of flight hardware. We use a cutoff of ± 1 mag from the Main Sequence to establish Main sequence character (see Fig. 3). The DASSC catalogue contains 1229 single main sequence stars of which 107 are F, 235 are G, 536 are K, and 351 are M type (incl. 14 M stars with no B-V index, for these stars the calculations are based on the V-K index).

The DASSC target star list was used for the trade off studies of different mission architectures for the proposed Darwin mission (see e.g. Kaltenegger and Karlsson 2004; Kaltenegger et al. 2006). Similar studies were done for TPF-I (see e.g. Lay et al. 2005; Dubovitsky and Lay 2004) and TPF-C (Brown 2006).

---

[2] The full DASSC catalogue can be obtained from one of the authors (LK) by sending an e-mail to lkaltene@cfa.harvard.edu or at http://cfa-www.harvard.edu/~lkaltenegger

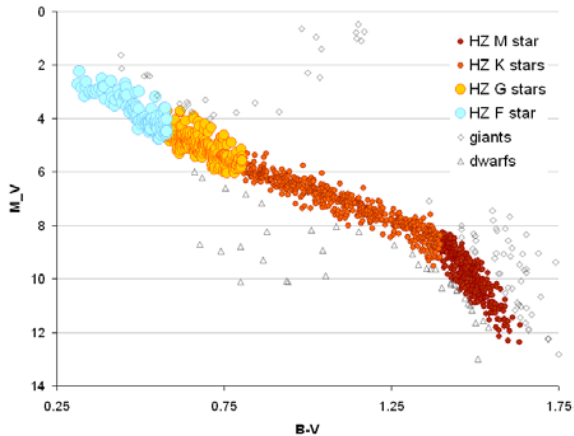

**Fig. 3 Color-magnitude diagram for the Single stars of the Darwin All Sky Target Stars Catalogue (DASSC).**

**2.1 Multiplicity**

Unsurprisingly, many of the preliminary targets are either spectroscopic binaries or have known resolved companions. Multiple stars are potentially interesting targets for searches for terrestrial exoplanets, since various groups have determined that stable orbits exist around multiple star systems. Further, Raghavan et al. (2006) found that about 30% of Extrasolar Giant Planets (EGP) host stars are multiple. It still remains to be investigated exactly what are the constraints under which a given binary can or cannot be observed. A second star in the vicinity, especially within the field of view, induces a high background signal that inhibits the detection of an orbiting planet.

All known multiple systems have been removed from the DASSC in the first determination of instrument-depending sub-catalogues. We note that 949 out of the 2303 all sky target stars are members of a double or multiple system, 716 tagged by CCDM. 211 additional multiple systems are flagged by Hipparcos, an additional 22 as spectroscopic binaries in SB9 and 125 are not main sequence stars. The list of multiple systems in the DASSC is far from complete. Also companion stars found outside the primary field of view need to be modeled in order to evaluate the effect of their light on the background level. Not all multiple systems will degrade the performance of the instrument. The apparent separation at the time of the observation will be important to determine the suitability of the multiple systems as a target.

**2.2 Sub-catalogues Derived for Specific Architectures**

An instrument that characterizes terrestrial planets (e.g. Darwin and Terrestrial Planet Finder) will operate best for nearby stars.

Occultation and coronagraph designs tend to have a set inner working angle (IWA) based on their mask design favoring a wider separation between planet and host star. The IWA translates into a minimum separation needed between a planet and its host star in order for the former to be detectable. In principle an interferometer searching for Earth-like planets is not restricted by distance. The maximum separation between the individual telescopes of an interferometer design sets the maximum resolution and can be adapted to each individual target system, thus detecting and characterizing planets very close to their stars. However, although the baseline can be adjusted for the required resolution, the integration time will at some distance be unrealistic for a given collecting area. An Earth size planet, located at a distance of 10 pc will have a flux of 0.34 µJy at 10 µm.

The stellar type of the host star influences the detectabilty of a terrestrial planet differently, depending on the wavelength of observation. To observe Earth-like planets in the HZ around a given star, the thermal flux will to first order be constant for a given planetary size, while the reflected stellar flux will scale with the brightness of the star and the distance of the HZ. The primary's thermal emission will, on the other hand, be progressively smaller for later and later spectral types (due to decreased temperature and area of the star), making to first order the detection of planets around cool stars easier at infrared wavelengths. The wider HZ around hot stars makes them better targets for optical wavelengths due to the IWA of coronagraph design.

Additional selection criteria will very likely be added because of constraints caused by the design and actual flight configuration of the instrument. Depending on the specific design, different parts of the sky will be accessible and sub-catalogues need to be derived for these. The actual target star list of a given subcatalogue can also influence the choice of technical implementation, e.g. all G stars within 30pc.

As an illustrative example we compare two designs of mission characterizing extrasolar planets, the current interferometer design for the Darwin study (the so-called 'EMMA' architecture) as well as one sub-catalogue that is based on an IWA of 75 mas representing optical designs. The sample is significantly smaller for the design with an IWA because the number of stars where the HZ is inaccessible rises rapidly both with distance and with the intrinsic faintness of the later type stars. Close as well as bright stars are the best targets if the design has an IWA.

### 2.2.1 EMMA Design: Sub-Catalogue

The EMMA interferometry configuration (Karlsson et al. 2003) is a design of three 3.5m-telescopes that fly in formation, with the beam combiner is flying at a large distance away from the plane of the light collecting telescopes, allow a pointing circle of ± 71.8° around the anti-solar point. These technical specifications lead to a subset of DASSC stars, forming this particular target list,

• 1178 single stars: 330 M stars, 514 K stars, 217 G stars and 103 F stars and 14 stars without B-V index

### 2.2.2 Effect of Inner Working Angles

This selection shows how an IWA influences the star sample. 75 mas was picked as a realistic value for a coronagraph design, 50 mas was selected to demonstrate the selection effect of an IWA on the target star class. We used the equivalent Sun-Earth separation as the cutoff for the stars in this sub-catalogue. If the whole HZ (including the inner edge of the HZ) should be probed, that would require an IWA smaller than 75 mas or 50 mas as shown here

• 89 single stars: 3 M stars, 12 K stars, 21 G stars and 53 F stars.

An Inner working angle of 50 mas leads to a much larger sample of target stars.

• 276 single stars: 21 M stars, 48 K stars, 111 G stars and 96 F stars.

## 3. Stellar Activity and Implications for Habitability

### 3.1 The Sun and its influence on Earth

Coordinated studies of extreme space weather events which effect the behavior of the upper atmosphere, ionosphere, magnetospheric environment and thermal and non-thermal atmospheric loss processes of Earth can serve as a proxy for the influence of the active young Sun or other stars with implications for the evolution of planetary atmospheres (solar system and exoplanets), water inventories and habitability. The most important solar processes that influence the space weather are: 1) coronal mass ejections (CMEs), huge bubbles of gas that are expelled from the Sun; 2) Flares, eruptive events where radiation plus energetic particles are released, and 3) Solar wind, a continuous stream of charged particles.

On Earth, there are two protection mechanisms: the atmosphere provides a shield against short wavelength radiation that is harmful for advanced life. UV radiation is mainly absorbed in the terrestrial ozone layer at heights between 12 and 50 km, X-rays are absorbed even higher in the atmosphere by molecular oxygen. Earth's magnetosphere provides a shield against charged particles, most of them are deflected, some can enter via reconnection processes in the magnetotail near the magnetic poles causing aurora. The solar radiation is variable by about 0.1 % over an eleven year solar activity cycle. The variation strongly depends on the wavelength, the shorter the wavelength, the stronger the variation (in the visible below 0.1%, in the UV more than 10 %.). Increased shortwave radiation from the Sun causes the higher atmospheric layers to expand, and also forms NO in the higher atmosphere what destroys part of the ozone.

At present, energetic events on the Sun are not likely to destroy habitability on our planet. The early Sun however was more active and therefore the effects of flares and CMEs and solar wind were considerably stronger on early life on Earth (Hanslmeier, 2007, 2009). The evolution of planetary atmospheres and their water inventories is strongly related to the evolution of the radiation and plasma outflow of their host stars which are much more intense compared to similar stars which are several billion years old, when the young stars arrive at the Zero-Age-Main-Sequence (Ayres 1997; Ribas et al., 2005; Wood et al., 2002; 2005; Güdel 2007).

**3.2 Influence of Stellar Activity on Planetary Atmospheres**

The closer the HZ is to the host star, the more effective the influence of mass ejecta and high energy radiation outbreaks is on its atmosphere. Especially at younger ages, stars emit high levels of radiation in short-wavelength ranges (X-ray, EUV, FUV, and UV). The activity phase of high mass stars is shorter than for low mass stars. During this time, enhanced flaring activity is also present, as well as stellar wind and coronal mass ejections. These phenomena could endanger a planet's atmosphere and/or the evolution of life, High-energy radiation heats the upper atmosphere of a planet and leads to enhanced atmospheric losses. Winds and CME-impacts can compress the magnetosphere of planets and lead to atmospheric erosion (Khodachenko et al. 2007). The difference in the mass flux of weak and strong CMEs is not as important as the difference of weakly and strongly magnetized terrestrial exoplanets. Weakly magnetized Earth-like exoplanets in close-in habitable zones - e.g. orbiting M stars - which are exposed to a high level of XUV radiation can potentially lose their entire atmosphere if exposed to CME plasma flow. Planets with a strong magnetic field and a high $CO_2$ mixing ratio can sustain the erosion due to CMEs if XUV fluxes are less than 50-70 times of the present solar XUV radiation (Lammer et al., 2007).

The activity of a star decreases with increasing age, but the timescales for this decrease depend on stellar mass with lower-mass stars sustaining a higher level of activity for longer timescales. Active stars exhibit frequent and powerful flares (Audard et al. 2000) that scale with the quiescent X-ray luminosity. The fraction of X-ray radiation that is emitted seems to have an upper limit of $\log(L_X/L_{bol}) \approx -3$ at which the activity is saturated (e.g. Vilhu & Walter 1987, Fleming et al. 1993, Pizzolato et al. 2003). The evolution of the XUV flux of solar analogs of different ages show that the flux between 0.1-120 nm scales as $t^{-1.23}$ (Ribas et al. 2005). Similar study are being currently carried out for M dwarfs (Guinan & Engle, 2009, Penz & Micela 2008), the Pleiades and Hyades (Penz et al. 2008) (Fig.4).

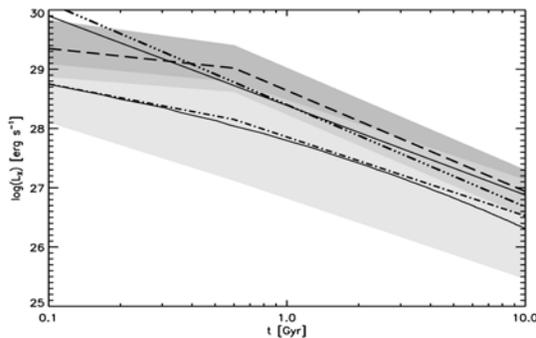

Fig. 4: Temporal evolution of the stellar X-ray luminosity. The dashed line gives the median X-ray luminosity of G stars, the dark shaded area the 1σ equivalent of the luminosity distributions, as derived by Penz et al. (2008). The dash-dotted line and the light shaded area show the same for M dwarfs (Penz & Micela, 2008). The solid lines display (from top to bottom) the scaling law from Ribas et al. (2005) for solar analogs in the range 0.1-10 nm, and the scaling from Guinan & Engle (2009) for a sample of M dwarfs, respectively. The dash-dot-dotted line displays the scaling law for G stars from Scalo et al. (2007).

The solar wind leads to a mass-loss rate of approximately $2 \cdot 10^{-14}$ $M_{sun}$ $yr^{-1}$. Several indirect methods were able to give first estimates on the mass-loss rates of solar-like and late-type stars (Wood et al. 2002, 2005, Wargelin & Drake 2001, 2002) between <0.2 and 100 times solar respectively. Observations indicate the existence of stellar analogues to solar CMEs and prominences (Jardine et al., 2008, Mullan et al. 1989, Bond et al. 2001). Currently, the stellar samples which can be used for stellar wind observations are not large enough to precisely address these questions. Further observations are required to complete the observational data base to characterize of the whole target star sample.

## 4. Conclusions

We discussed the criteria for selecting potential target stars suitable for the search for Earth like planets, with a special emphasis on the stellar aspects of habitability. Missions that search for terrestrial exoplanets will explore the presence and habitability of Earth-like exoplanets around several hundred nearby stars, mainly F, G, K, and M stars. Using the Darwin All Sky Star Catalogue (DASSC), we discuss the influence of different designs and derive configuration dependent sub-catalogues.

The detection of several terrestrial exoplanets in the HZ of stars will allow us for the first time to carry out real comparative planetology, relating planet properties like masses, sizes, orbits, and atmospheric physics to the astrophysical characteristics of their host stars.

**Acknowledgements**

LK gratefully acknowledges support from the Harvard Origins of Life Initiative. We thank ISSI. This research has made use of the SIMBAD database, operated at CDS, Strasbourg, France, NASA's Astrophysics Data System and data products from the Two Micron All Sky Survey, a joint project of the UMass and IPAC, funded by NASA and NSF.

Kaltenegger et al., Astrobiology 2009Super-Earth Orbiting HD 7924, arXiv0901.4394H

Jardine, M., Donati, J., and Gregory, S. G. (2008) Magnetic coronae of active main-sequence stars. arXiv0811.1906.

Joshi, M.M. (2003) Climate Model Studies of Synchronously Rotating Planets. Astrobiology, 3, 2:415-427.

Joshi, M.M., Haberle, R.M., and Reynolds, R.T. (1997) Simulations of the Atmospheres of Synchronously Rotating Terrestrial Planets Orbiting M Dwarfs: Conditions for Atmospheric Collapse and the Implications for Habitability, Icarus, 129, 2: 450-465.

Kaltenegger, L., Eiroa, C., and Fridlund, M. (2008) Target star catalog for Darwin: Nearby Stellar sample for a search for terrestrial planets, arXiv0810.5138K

Kaltenegger, L, Jucks, K. W., and Traub, W. A. (2007) Spectral Evolution of an Earth-like Planet, ApJ, 658: 598-616.

Kaltenegger, L., and Karlsson, A.L. (2004) Requirements on the stellar rejection for the Darwin Mission. In New Frontiers in Stellar Interferometry, Proceedings of SPIE Volume 5491. edited by W.A. Traub. Bellingham, WA, The International Society for Optical Engineering, pp 275, 275

Kaltenegger, L. (2004) Search for Extra-Terrestrial planets: The DARWIN mission - Target Stars and Array Architectures (PhD thesis), 2005astro.ph..4497K

Kasting, J.F., Whitmire, D.P., and Reynolds, R.T. (1993) Habitable Zones around Main Sequence Stars, Icarus 101, 1:108-128.

Khodachenko, M. L., et al. (2007) Coronal Mass Ejection (CME) Activity of Low Mass M Stars as an Important Factor for the Habitability of Terrestrial Exoplanets. I. CME Impact on Expected Magnetospheres of Earth-Like Exoplanets in Close-In Habitable Zones. Astrobiology 7, 167-184.

Kjeldsen, H., Bedding, T.R., and Christensen-Dalsgaard, J. (2009) Measurements of Stellar Properties through Asteroseismology. In A Tool for Planet Transit Studies. IAU Symposium 253 on Transiting Planets, pp 309-317.

Lammer, H., et al. (2007) Coronal Mass Ejection (CME) Activity of Low Mass M Stars as an Important Factor for the Habitability of Terrestrial Exoplanets. II. CME-Induced Ion Pick Up of Earth-Like Exoplanets in Close-In Habitable Zones. Astrobiology 7, 185-207.

Lay, O.P., et al. (2005) Architecture trade study for the Terrestrial Planet Finder Interferometer. In: Techniques and Instrumentation for Detection of Exoplanets II. edited by Coulter, Daniel R. Proceedings of the SPIE, Volume 5905, pp. 8-20.

Léger, A. et al. (2009) Transiting exoplanets from the CoRoT space mission. VIII. CoRoT-7b: the first super-Earth with measured radius, A&A,506, 1, pp.287-302

Lunine, J.I. (2001) Special Feature: The occurrence of Jovian planets and the habitability of planetary systems. Publications of the National Academy of Science, 98:809-814.

Mamajek, E. E., and Hillenbrand, L. A. (2008) Improved Age Estimation for Solar-Type Dwarfs Using Activity-Rotation Diagnostics. ApJ, 687:1264-1293.

Mamajek, E.E., Barrado y Navascués, D., Randich, S., Jensen, E.L.N., Young, P.A., Miglio, A., and Barnes, S.A. (2008) A Splinter Session on the Thorny Problem of Stellar Ages. In 14th Cambridge Workshop on Cool Stars, Stellar Systems, and the Sun, edited by G. van Belle, ASP Conference Series, Vol. 384, 374-382.

Mayor, M., et al. (2009) The HARPS search for southern extra-solar planets. XIII. A planetary system with 3 super-Earths (4.2, 6.9, and 9.2 M$_\oplus$), A&A, 493, 2:639-644.

Mullan, D. J., Sion, E. M., Bruhweiler, F. C., and Carpenter, K. G. (1989) Evidence for a cool wind from the K2 dwarf in the detached binary V471 Tauri. ApJ. 339:L33-L36.

Penz, T., and Micela, G. (2008) X-ray induced mass loss effects on exoplanets orbiting dM stars. A&A. 479: 579-584.

Penz, T., Micela, G., and Lammer, H. (2008) Influence of the evolving stellar X-ray luminosity distribution on exoplanetary mass loss. A&A. 477:309-314.

Perryman, M.A.C., et al. (2003) The stellar activity-rotation relationship revisited:12